\documentclass[conference]{IEEEtran}
\IEEEoverridecommandlockouts
\usepackage{cite}
\usepackage{amsmath,amssymb,amsfonts}
\usepackage{algorithmic}
\usepackage{graphicx}
\usepackage{textcomp}
\usepackage{xcolor}
\def\BibTeX{{\rm B\kern-.05em{\sc i\kern-.025em b}\kern-.08em
    T\kern-.1667em\lower.7ex\hbox{E}\kern-.125emX}}
\usepackage{hyperref}
\hypersetup{%
  colorlinks=true,%
  linkcolor={red!50!black},
  citecolor={blue!65!black},
  urlcolor={blue!80!black},
  bookmarksnumbered=true,
  bookmarksopen=true}
\usepackage{makecell}
\usepackage{booktabs}
\usepackage{multirow}
\newcommand\NameOrcid[2]{\hypersetup{hidelinks}\href{https://orcid.org/#2}{#1}}

\begin{document}

\title{\huge Exploring LLM-based Verilog Code Generation with Data-Efficient Fine-Tuning and Testbench Automation
\thanks{This work has been submitted to the IEEE for possible publication. Copyright may be transferred without notice, after which this version may no longer be accessible.}
\thanks{We acknowledge the financial support from Academia Sinica's SiliconMind Project (AS-IAIA-114-M11). This work was also supported in part by National Science and Technology Council, Taiwan, under Grants 112-2221-E-002 -159 -MY3 and 114-2221-E-006 -165 -MY3. We also thank National Center for High-performance Computing (NCHC) and Taipei-1 for providing computational and storage resources.}
}


\author{
    \IEEEauthorblockN{\NameOrcid{Mu-Chi Chen}{0009-0007-6013-4122}\IEEEauthorrefmark{1}, \NameOrcid{Po-Hsuan Huang}{0000-0002-7458-9634}\IEEEauthorrefmark{2}, \NameOrcid{Yu-Hung Kao}{0009-0002-6991-8795}\IEEEauthorrefmark{2}, \NameOrcid{Yen-Fu Liu}{0009-0003-1106-099X}\IEEEauthorrefmark{2}, \NameOrcid{Yu-Kai Hung}{0009-0007-0310-4883}\IEEEauthorrefmark{2},\\ \NameOrcid{Cheng Liang}{0009-0009-1532-3332}\IEEEauthorrefmark{2}, \NameOrcid{Shao-Chun Ho}{0009-0000-7363-2947}\IEEEauthorrefmark{2}, \NameOrcid{Chia-Heng Tu}{0000-0001-8967-1385}\IEEEauthorrefmark{3}, and \NameOrcid{Shih-Hao Hung}{0000-0003-2043-2663}\IEEEauthorrefmark{2}}\vspace{5pt}
    \IEEEauthorblockA{\IEEEauthorrefmark{1}Academia Sinica, Taipei, Taiwan}
    \IEEEauthorblockA{\IEEEauthorrefmark{2}National Taiwan University, Taipei, Taiwan}
    \IEEEauthorblockA{\IEEEauthorrefmark{3}National Cheng Kung University, Tainan, Taiwan}
}

\maketitle

\begin{abstract}
Recent advances in large language models have improved code generation, but their use in hardware description languages is still limited. Moreover, training data and testbenches for these models are often scarce. This paper presents a workflow that uses multi-agent models to generate testbenches for high-quality fine-tuning data. By automating testbench creation, the fine-tuned model for the specification-to-Verilog task achieves performance comparable to state-of-the-art methods on the refined VerilogEval v2 benchmark while using less training data. This study provides a basis for future work on LLM-based HDL generation and automated verification.
\end{abstract}


\section{Introduction}

The design and verification of digital circuits rely on hardware description languages (HDLs) such as Verilog, which allow engineers to describe and simulate hardware behavior at multiple abstraction levels~\cite{designs4030031}. As circuit complexity increases, manual coding and verification have become time-consuming and error-prone, creating a demand for automated tools that support efficient module generation and testing. Recent progress in large language models has shown strong capabilities in code understanding and generation~\cite{GPT3,GPT4,Guo2025}, indicating potential for HDL-related applications.

Despite these advances, the use of large language models (LLMs) in hardware design remains limited. HDLs have structural and behavioral properties that differ from conventional programming languages, which makes them more difficult for language models to handle. Limited access to advanced proprietary LLMs further restricts research, customization, and reproducibility, especially in settings where design information must remain confidential. Dataset and its quality are also significant concerns, as the lack of diverse and representative training data can hinder model performance. These constraints highlight the need for accessible and domain-adapted solutions in hardware design automation.

This study introduces a workflow that uses LLMs to generate testbenches and create high-quality data for fine-tuning Verilog generation models. The workflow adopts a multi-agent structure in which one agent produces Verilog modules based on design specifications, and another generates testbenches for verification. This division improves code accuracy and enables automatic production of training data by integrating LLM agents with existing verification tools.

Our experimental results demonstrate that the proposed LLM-based multi-agent workflow can generate syntactically correct and functionally relevant Verilog code as training data. The approach also accelerates testbench creation, improving verification coverage and reducing manual effort. With significantly reduced training data compared to existing state-of-the-art methods, we show that LLMs can effectively contribute to HDL design and verification tasks through systematic evaluation on benchmarks. This research lays the foundation for future work in applying LLMs to electronic design automation (EDA), opening new opportunities for AI-assisted hardware development and intelligent verification systems. The main contributions of this paper are summarized as follows:
\begin{itemize}
  \item We propose a novel workflow to evaluate two data generation strategies for fine-tuning LLMs on specification-to-Verilog code tasks.
  \item We present a workflow that uses multi-agent LLMs for Verilog generation and automated testbench creation, improving data quality and efficiency.
  \item We evaluate the approach on the refined VerilogEval v2 benchmark and show that it produces functionally accurate Verilog code while achieving performance comparable to state-of-the-art methods.
\end{itemize}

\section{Background}\label{sec:Background}
This section presents an overview of Verilog, testbenches, and LLMs. It also reviews related research in the field of LLM-based code generation and verification to establish the foundation for this study.

\subsection{Verilog and Testbenches}

Verilog is a foundational hardware description language that supports combinational and sequential logic across multiple abstraction levels. It is widely used to model circuit behavior from register-transfer level descriptions to gate-level implementations, and its flexibility supports efficient prototyping with standard simulation and synthesis tools \cite{electronics12183821}.  
Verifying Verilog modules is essential because design errors can propagate through later development stages. Tools such as AVERT \cite{9826162} automate testbench generation and reduce manual effort for combinational and sequential designs. For complex systems, the Universal Verification Methodology with coverage-driven verification remains a practical and effective approach \cite{electronics12183821}.

\subsection{LLMs for Hardware Design}

LLMs represent a major advancement in artificial intelligence and can process text, perform reasoning, and generate code across many programming languages~\cite{GPT3,GPT4}. Although these models handle syntax and semantics well in languages such as Python and C++, applying them to hardware design remains challenging. HDLs differ from software languages because they describe concurrent operations, timing behavior, and signal interactions under strict design rules. These characteristics require domain-specific training and specialized workflows for effective LLM use in hardware design.

Leveraging LLMs to assist in or fully automate testbench generation offers a promising pathway toward intelligent verification workflows.
Recent work has extended verification methodologies to machine learning-based approaches. For instance, Wang et al.~\cite{VeriPrefer} integrated testbench feedback into the reinforcement learning process to improve the functional correctness of LLMs generating Verilog code. Similarly, VeriReason~\cite{VeriReason} enhances reasoning in Verilog generation using reinforcement learning with testbench feedback. 

Zhu et al.~\cite{zhu2025qimengcodevr1reasoningenhancedveriloggeneration} introduced CodeV-R1, which combines rule-based testbench generation and round-trip data synthesis for reliable Verilog generation from natural language descriptions. They employ both SFT and RL (DAPO~\cite{yu2025dapoopensourcellmreinforcement}) to enhance Verilog coding capabilities. The key part is that CodeV-R1 enhances base LLMs with reasoning capabilities by distilling DeepSeek-R1~\cite{Guo2025}'s reasoning traces and outputs during its supervised fine-tuning (SFT) stage. CodeV-R1's methodology significantly increases the accuracy of 7B models, enabling them to exhibit R1-style reasoning capabilities for solving challenging Verilog problems. From CodeV-R1's results, we found that 7B models can also achieve good RTL coding capability after being equipped with reasoning capabilities via distillation from DeepSeek-R1 in the SFT stage.

\section{Methodology}\label{sec:Methodology}

\begin{figure}[htbp]
  \centering
  \includegraphics[width=0.8\linewidth]{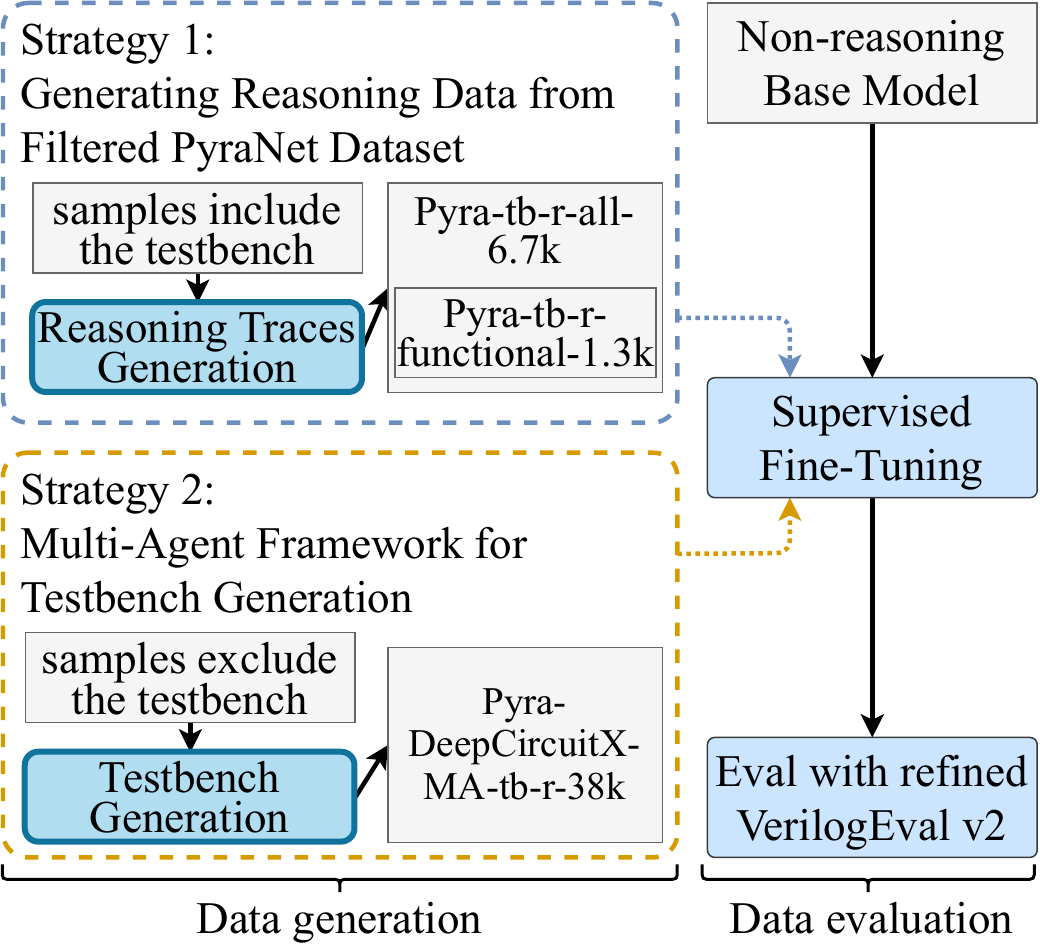}
  \caption{Proposed workflow for Verilog code training data generation.}
  \label{fig:workflow}
  \vspace{-10pt}
\end{figure}

Our proposed workflow consists of several key components that leverage LLMs for Verilog code generation and testing. As shown in \figurename~\ref{fig:workflow}, the workflow begins with generating reasoning data from a filtered PyraNet dataset using DeepSeek-R1. This reasoning data serves as high-quality training input for SFT of base LLMs, enhancing their Verilog coding capabilities. Another strategy in our workflow is the multi-agent framework for automated testbench generation. This framework employs multiple LLM agents, each specializing in different aspects of testbench creation, to collaboratively produce comprehensive verification environments. After data generation, an SFT process is applied to fine-tune base LLMs to evaluate their performance on Verilog code generation tasks.

\subsection{Generating Reasoning Data from Filtered PyraNet Dataset}

To train reasoning models, we first needed reasoning traces. However, datasets typically provide only design specifications and golden solutions (code). To address this, we used vLLM to deploy DeepSeek-R1 on 16x H100 GPUs to generate reasoning traces. We selected the Pyra-tb dataset (6.7k samples), filtered from 692k~\cite{Nadimi_2025}, for this purpose. This dataset is considered the highest-quality subset produced by VeriPrefer~\cite{VeriPrefer}, following the three filtering process. Notably, the fourth block includes testbenches, which can be used to verify the functional correctness of the generated Verilog code.

\subsection{Multi-Agent Framework for Testbench Generation}

To automate the creation of testbenches for the Verilog code, we propose a multi-agent framework that leverages the capabilities of LLMs. In this framework, multiple agents are assigned specific roles in the testbench generation process, allowing for a collaborative approach to creating comprehensive reasoning traces and corresponding testbenches. Each agent is responsible for a distinct aspect of tasks, such as specification quality check and testbench generation. The agents communicate and collaborate to ensure that the generated testbenches are thorough and cover a wide range of scenarios. This multi-agent approach not only enhances the quality of the testbenches but also improves the efficiency of the generation process, as each agent can focus on its area of expertise.

\begin{figure}[htbp]
  \centering
  \includegraphics[width=.95\linewidth]{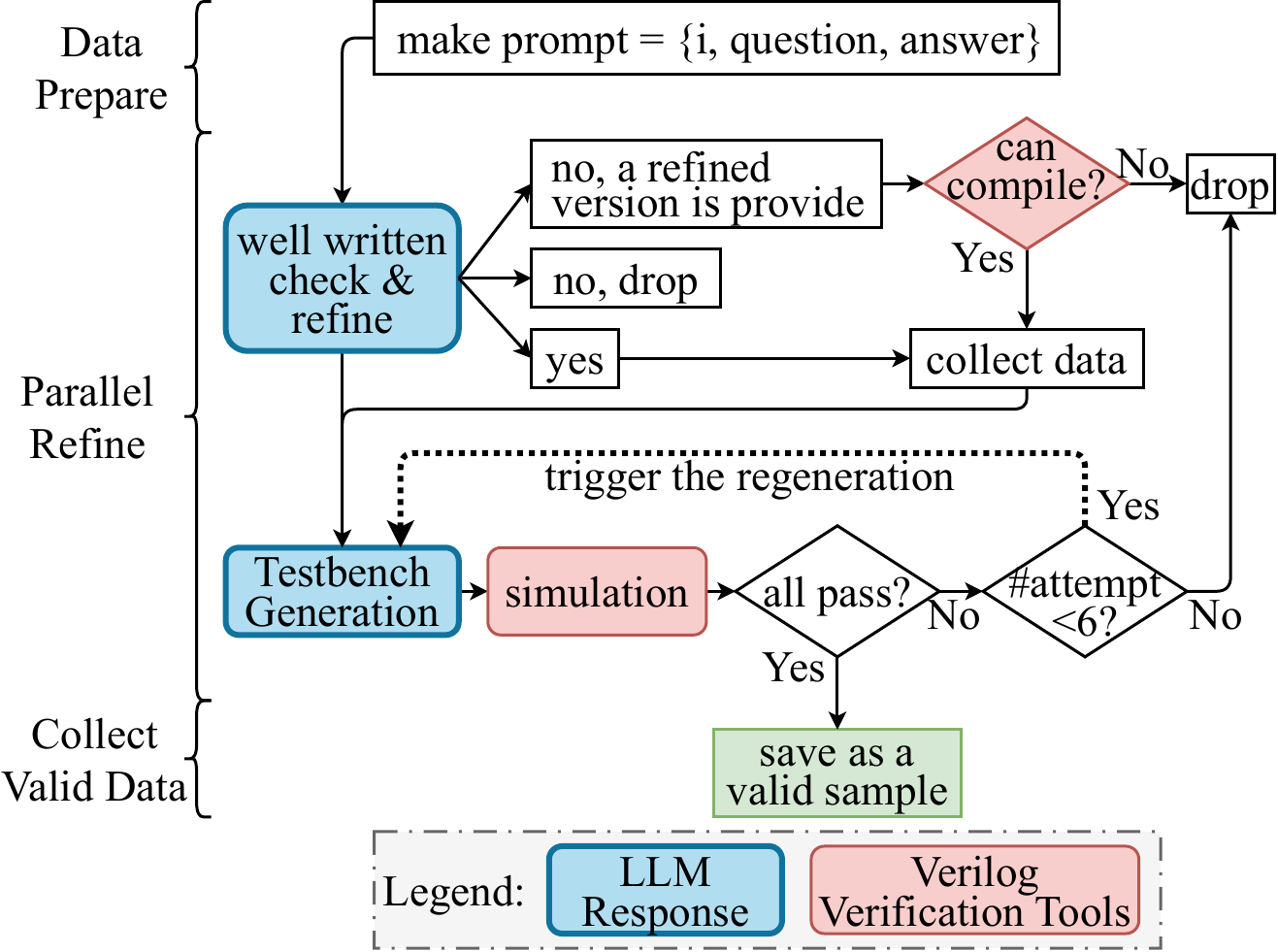}
  \setlength{\abovecaptionskip}{0pt}
  \caption{Multi-agent framework for testbench generation.}
  \label{fig:tb_gen}
  \vspace{-10pt}
\end{figure}

As shown in \figurename~\ref{fig:tb_gen}, our multi-agent framework consists of the following key components: 1) well written check and refine; 2) testbench generation; 3) verilog verification tools (compiler and simulator).
In the pipeline, we use two datasets, PyraNet~\cite{Nadimi_2025} and DeepCircuitX~\cite{11105939} as the input of the multi-agent framework. The input code is first sent to the quality check agent, which tries to refine the question description if the initial description is unclear or incomplete. This description is then passed to the testbench generation agent, which creates a comprehensive testbench that includes the necessary components for simulation and verification. Finally, the generated testbench is reviewed and revised to ensure its correctness.

\begin{figure}[htbp]
  \centering
  \includegraphics[width=.95\linewidth]{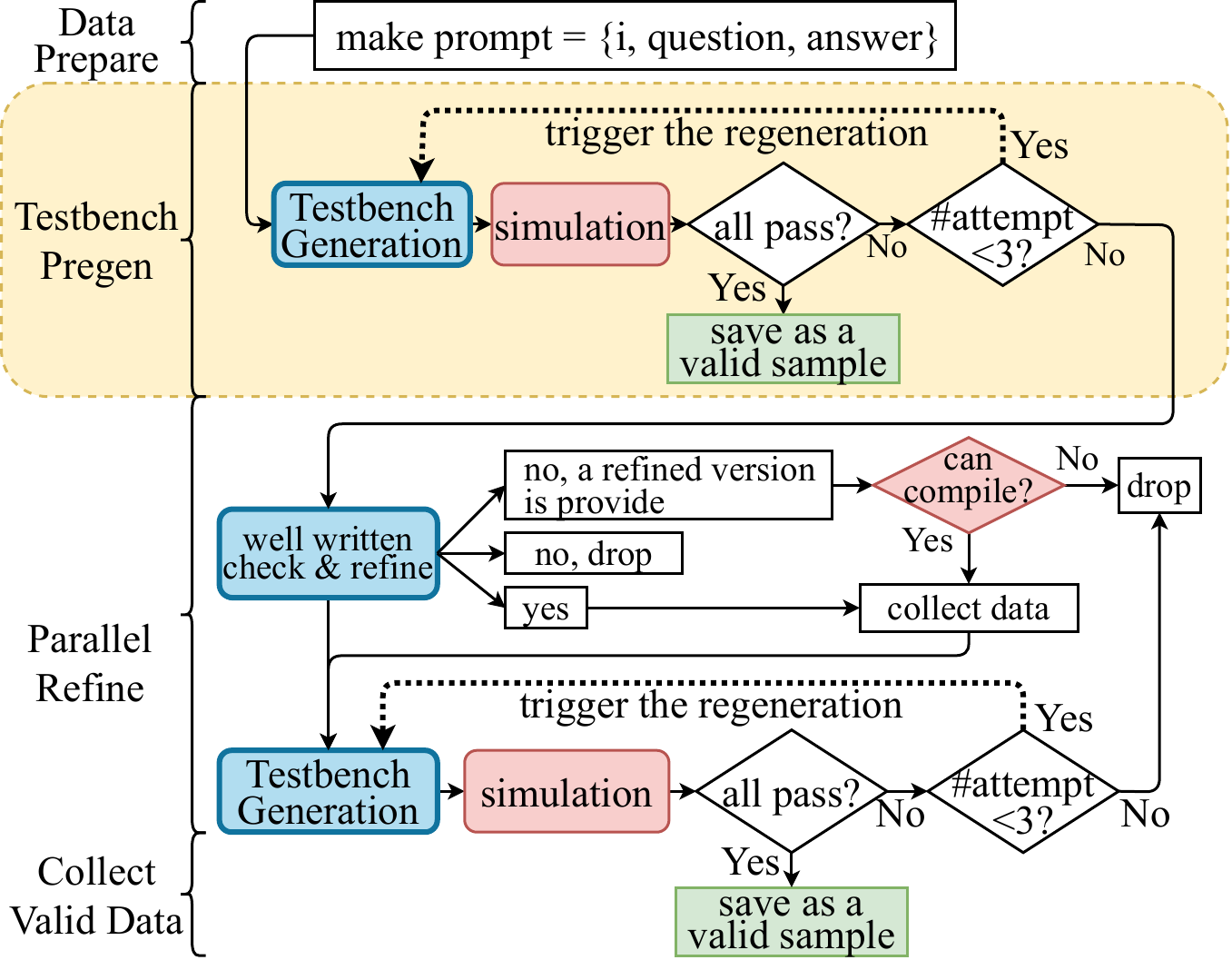}
  \setlength{\abovecaptionskip}{0pt}
  \caption{Multi-agent framework with pregenerating testbench.}
  \label{fig:tb_gen_mix}
  \vspace{-10pt}
\end{figure}

To enhance the efficiency of the testbench generation process, we propose a variant of the multi-agent framework that incorporates pre-generated testbenches, as illustrated in \figurename~\ref{fig:tb_gen_mix}. In this approach, testbenches are generated prior to the application of the well written check and refine agent. The workflow includes two stages of testbench revision (one before and one after question refinement) which leads to a higher pass rate compared to the original framework. This modification not only improves the overall quality of the generated testbenches but also reduces the time and manual effort required to produce reliable verification environments.

\section{Evaluation}\label{sec:Evaluation}

To evaluate the effectiveness of our proposed workflow, we conducted a series of experiments using a set of benchmark Verilog modules. 
A base model, Qwen-Coder-7B-Instruct~\cite{Qwen}, was evaluated.
We measured the performance of the LLM in generating syntactically correct and functionally relevant Verilog code. Additionally, we evaluated the quality of the automatically generated testbenches by assessing their ability to detect functional errors in the generated code.

\subsection{Experimental Setup}

Almost all experiments were conducted on Nvidia DGX H100 clusters, except for agents (GPT-4.1), which ran on remote cloud instances. The hardware and software specifications are listed in Table~\ref{tab:exp_setup}.

\begin{table}[h]
  \vspace{-10pt}
  \centering
  \caption{Experimental Setup.}
  \label{tab:exp_setup}
  \begin{tabular}{ll}
    \toprule
    Component & Specification \\
    \midrule
    Hardware & Nvidia DGX H100 Cluster (2-node) \\
    GPU & 8x Nvidia H100 GPUs per node \\
    CPU & 2x Intel Xeon Platinum 8480+ per node \\
    Software & RHEL 8.10 \\
    Python Version & Python v3.12.9 \\
    Packages & \makecell[l]{vLLM v0.10.1.1, PyTorch v2.7.1,\\trl v0.21.0, transformers v4.55.4} \\
    Verilog Compiler & Icarus Verilog v11.0 \\
    \bottomrule
  \end{tabular}
  \vspace{-10pt}
\end{table}

\subsection{Refinement of VerilogEval v2}

Our experiments use a refined version of VerilogEval v2, a benchmark for evaluating LLMs' Verilog code generation. VerilogEval v2~\cite{pinckney2025revisitingverilogevalyearimprovements} contains coding problems with specifications and test cases to validate generated code.  
We refined the benchmark by updating test cases to cover more scenarios and edge cases, enhancing specifications with detailed requirements, and addressing ambiguities from initial experiments. These improvements provide a clearer and more comprehensive assessment of LLM performance.  
The refinements led to notable performance gains; for example, CodeV-R1-7B-Distill rose from 65\% to 70\%, and CodeV-R1-7B (w/ DAPO) from 69\% to 74\%.

\subsection{Generating Reasoning Data from Filtered PyraNet Dataset}

A total of 6,704 samples from the Pyra-tb dataset (each including a corresponding testbench generated by VeriPrefer\cite{VeriPrefer}'s data processing flow) were processed on the DGX H100 cluster. Using DeepSeek-R1, reasoning traces were generated for each sample, resulting in 1,386 functionally correct samples, 4,421 syntactically correct samples, and 897 compilation errors. The entire run took about 55 hours. In total, 6,073,695 input tokens and 54,778,748 output tokens were processed, achieving an average throughput of 305.225 tokens per second.

\subsection{Generating Testbenches with Multi-Agent Framework}

We applied our multi-agent framework to generate testbenches for Verilog code from PyraNet~\cite{Nadimi_2025} and DeepCircuitX~\cite{11105939}. The framework successfully produced testbenches capable of validating the functionality of the generated Verilog code. We evaluated their quality by measuring coverage and effectiveness in detecting functional errors. The results demonstrate that the multi-agent framework can generate high-quality testbenches that improve the verification process.
Tables~\ref{tab:tb_pyranet} and~\ref{tab:tb_deepcircuitx} present statistics for the testbenches generated for the PyraNet and DeepCircuitX datasets, respectively. For each dataset, we tested 50 questions using GPT-4.1 and averaged the results over three runs. In a run, stages were executed from top to bottom; for example, an average of 4.33 questions passed at the first stage with pregeneration, and all stages yielded 38 passes out of 50. The findings show that incorporating pregenerated testbenches increases the total number of passes while reducing API call counts, highlighting the efficiency and effectiveness of our framework, particularly on DeepCircuitX.

\begin{table}[h]
  \vspace{-10pt}
  \caption{Multi-agent testbench generation statistics on PyraNet (average over 3 runs, each run process 50 questions).}
  \label{tab:tb_pyranet}
  \centering
  \begin{tabular}{lrr}
    \toprule
    Stage & \makecell{w/o pregenerating \\ testbenches} & \makecell{w/ pregenerating \\ testbenches} \\
    \midrule
    Init testbench pass & 13.33 & 4.33 \\
    1st retry pass & 13.00 & 13.00 \\
    2nd retry pass & 3.33 & 10.33 \\
    3rd retry pass & 1.33 & 2.67 \\
    Revision init & -- & 2.00 \\
    4th retry pass & 0.00 & 3.67 \\
    5th retry pass & 0.00 & 1.33 \\
    6th retry pass & 0.33 & 0.67 \\
    \midrule
    Total \#Pass $\uparrow$ & 31.33 & \textbf{38.00} \\
    API Count $\downarrow$ & 237.67 & \textbf{234.33} \\
    \bottomrule
  \end{tabular}
  \vspace{-10pt}
\end{table}

\begin{table}[h]
  \caption{Multi-agent testbench generation statistics on DeepCircuitX (average over 3 runs, each run processes 50 questions).}
  \label{tab:tb_deepcircuitx}
  \centering
  \begin{tabular}{lrr}
    \toprule
    Stage & \makecell{w/o pregenerating \\ testbenches} & \makecell{w/ pregenerating \\ testbenches} \\
    \midrule
    Init testbench pass & 6.67 & 6.33 \\
    1st retry pass & 9.33 & 14.00 \\
    2nd retry pass & 3.33 & 4.67 \\
    3rd retry pass & 1.33 & 3.33 \\
    Revision init & -- & 0.67 \\
    4th retry pass & 0.33 & 2.33 \\
    5th retry pass & 0.33 & 2.33 \\
    6th retry pass & 0.00 & 0.33 \\
    \midrule
    Total \#Pass $\uparrow$ & 21.33 & \textbf{34.00} \\
    API Count $\downarrow$ & 295.00 & \textbf{247.67} \\
    \bottomrule
  \end{tabular}
\end{table}

\subsection{Data Evaluation Results}

Table~\ref{tab:results} presents the performance of several models on the generated dataset. The results show that dataset size is critical for domain-specific tasks. The 1.3k high-quality reasoning samples did not improve performance, while the 6.7k samples with syntax errors increased accuracy to 47\%. Our fine-tuned model, MA-tb-7B, produced with the multi-agent generation pipeline, achieved a pass@1 rate of 68\%, which is competitive with state-of-the-art models while requiring less training data. This confirms the effectiveness of the multi-agent testbench generation framework and the supervised fine-tuning process for improving LLM performance in Verilog code generation.

\begin{table}[htbp]
  \centering
  \caption{Performance of Various Models on the Refined VerilogEval v2 Benchmark.}
  \setlength{\tabcolsep}{2.0pt}
  \label{tab:results}
  \begin{tabular}{llccc}
    \toprule
    Category & Model & \#Params & \makecell{\#Training\\data} & \makecell{pass@1\\(\%)} \\
    \midrule
    \multirow{3}{*}{\makecell[l]{Foundation \\ Model}}
    & Gemini2.5-Pro & n/a & n/a & 91 \\
    & DeepSeek-V3.1 & 671B & n/a & 74 \\
    & DeepSeek-R1-0528 & 685B & n/a & 81 \\
    \midrule
    \makecell[l]{Base \\ Model}
    & Qwen-Coder-7B-Instruct & \phantom{0}7B & n/a & 40 \\
    \midrule
    \multirow{2}{*}{\makecell[l]{Related \\ Work}}
    & CodeV-R1-7B-Distill & \phantom{0}7B & 87k & 70 \\
    & CodeV-R1-7B (w/ DAPO) & \phantom{0}7B & 87k{+}3.1k & 74 \\
    \midrule
    \multirow{3}{*}{Ours}
    & Pyra-tb-r-functional-7B & \phantom{0}7B & 1.3k & 40 \\
    & Pyra-tb-r-all-7B & \phantom{0}7B & 6.7k & 47 \\
    & MA-tb-7B & \phantom{0}7B & 38k & 68 \\
    \bottomrule
  \end{tabular}
  \vspace{-5pt}
\end{table}

\section{Conclusion}\label{sec:Conclusion}

This paper presents a workflow for AI-assisted hardware development and intelligent verification systems, where multi-agent LLMs generate training data for a non-reasoning base model performing spec-to-Verilog generation. By integrating automated testbench generation, the approach improves data productivity. Experiments show that the fine-tuned model produces correct and relevant Verilog code. This work lays a foundation for LLM-based HDL generation and verification, with potential impact on digital circuit design automation.

\bibliographystyle{IEEEtran}
\bibliography{paper}

@inproceedings{Nadimi_2025,
  title     = {{PyraNet}: A Multi-Layered Hierarchical Dataset for Verilog},
  doi       = {10.1109/dac63849.2025.11133406},
  booktitle = {2025 62nd ACM/IEEE Design Automation Conference (DAC)},
  publisher = {IEEE},
  author    = {Nadimi, Bardia and Boutaib, Ghali Omar and Zheng, Hao},
  year      = {2025},
  month     = jun,
  pages     = {1--7}
}

@inproceedings{11105939,
  author         = {Li, Zeju and Xu, Changran and Shi, Zhengyuan and Peng, Zedong and Liu, Yi and Zhou, Yunhao and Zhou, Lingfeng and Ma, Chengyu and Zhong, Jianyuan and Wang, Xi and Zhao, Jieru and Chu, Zhufei and Yang, Xiaoyan and Xu, Qiang},
  comment-author = {Li, Zeju and Xu, Changran and Shi, Zhengyuan and others},
  booktitle      = {2025 IEEE International Conference on LLM-Aided Design (ICLAD)},
  title          = {{DeepCircuitX}: A Comprehensive Repository-Level Dataset for RTL Code Understanding, Generation, and PPA Analysis},
  year           = {2025},
  volume         = {},
  number         = {},
  pages          = {204--211},
  doi            = {10.1109/ICLAD65226.2025.00029}
}

@inproceedings{GPT3,
  comment-author = {Brown, Tom B. and Mann, Benjamin and Ryder, Nick and Subbiah, Melanie and Kaplan, Jared and Dhariwal, Prafulla and Neelakantan, Arvind and Shyam, Pranav and Sastry, Girish and Askell, Amanda and Agarwal, Sandhini and Herbert-Voss, Ariel and Krueger, Gretchen and Henighan, Tom and Child, Rewon and Ramesh, Aditya and Ziegler, Daniel M. and Wu, Jeffrey and Winter, Clemens and Hesse, Christopher and Chen, Mark and Sigler, Eric and Litwin, Mateusz and Gray, Scott and Chess, Benjamin and Clark, Jack and Berner, Christopher and McCandlish, Sam and Radford, Alec and Sutskever, Ilya and Amodei, Dario},
  author         = {Brown, Tom B. and Mann, Benjamin and Ryder, Nick and others},
  title          = {Language models are few-shot learners},
  year           = {2020},
  comment-isbn   = {9781713829546},
  publisher      = {Curran Associates Inc.},
  address        = {Red Hook, NY, USA},
  booktitle      = {Proceedings of the 34th International Conference on Neural Information Processing Systems (NIPS)},
  articleno      = {159},
  numpages       = {25},
  location       = {Vancouver, BC, Canada},
  comment-series = {NIPS '20},
  comment-url    = {https://proceedings.neurips.cc/paper_files/paper/2020/file/1457c0d6bfcb4967418bfb8ac142f64a-Paper.pdf}
}

@inproceedings{9826162,
  author    = {McEllin, Jack and Conway, Richard and Ryan, Conor},
  booktitle = {2022 33rd Irish Signals and Systems Conference (ISSC)},
  title     = {{AVERT}: An Automatic Verilog Testbench Generation Tool for Grammatical Evolution},
  year      = {2022},
  volume    = {},
  number    = {},
  pages     = {1-8},
  doi       = {10.1109/ISSC55427.2022.9826162}
}

@inproceedings{zhu2025qimengcodevr1reasoningenhancedveriloggeneration,
  title          = {{QiMeng-CodeV-R1}: Reasoning-Enhanced Verilog Generation},
  author         = {Yaoyu Zhu and Di Huang and Hanqi Lyu and Xiaoyun Zhang and Chongxiao Li and Wenxuan Shi and Yutong Wu and Jianan Mu and Jinghua Wang and Yang Zhao and Pengwei Jin and Shuyao Cheng and Shengwen Liang and Xishan Zhang and Rui Zhang and Zidong Du and Qi Guo and Xing Hu and Yunji Chen},
  comment-author = {Zhu, Yaoyu and Huang, Di and Lyu, Hanqi and and others},
  booktitle      = {Proceedings of the 39th Annual Conference on Neural Information Processing Systems (NeurIPS)},
  year           = {2025},
  comment-url    = {https://arxiv.org/abs/2505.24183}
}

@article{Guo2025,
  comment-author = {Guo, Daya and Yang, Dejian and Zhang, Haowei and Song, Junxiao and Wang, Peiyi and Zhu, Qihao and Xu, Runxin and Zhang, Ruoyu and Ma, Shirong and Bi, Xiao and Zhang, Xiaokang and Yu, Xingkai and Wu, Yu and Wu, Z. F. and Gou, Zhibin and Shao, Zhihong and Li, Zhuoshu and Gao, Ziyi and Liu, Aixin and Xue, Bing and Wang, Bingxuan and Wu, Bochao and Feng, Bei and Lu, Chengda and Zhao, Chenggang and Deng, Chengqi and Ruan, Chong and Dai, Damai and Chen, Deli and Ji, Dongjie and Li, Erhang and Lin, Fangyun and Dai, Fucong and Luo, Fuli and Hao, Guangbo and Chen, Guanting and Li, Guowei and Zhang, H. and Xu, Hanwei and Ding, Honghui and Gao, Huazuo and Qu, Hui and Li, Hui and Guo, Jianzhong and Li, Jiashi and Chen, Jingchang and Yuan, Jingyang and Tu, Jinhao and Qiu, Junjie and Li, Junlong and Cai, J. L. and Ni, Jiaqi and Liang, Jian and Chen, Jin and Dong, Kai and Hu, Kai and You, Kaichao and Gao, Kaige and Guan, Kang and Huang, Kexin and Yu, Kuai and Wang, Lean and Zhang, Lecong and Zhao, Liang and Wang, Litong and Zhang, Liyue and Xu, Lei and Xia, Leyi and Zhang, Mingchuan and Zhang, Minghua and Tang, Minghui and Zhou, Mingxu and Li, Meng and Wang, Miaojun and Li, Mingming and Tian, Ning and Huang, Panpan and Zhang, Peng and Wang, Qiancheng and Chen, Qinyu and Du, Qiushi and Ge, Ruiqi and Zhang, Ruisong and Pan, Ruizhe and Wang, Runji and Chen, R. J. and Jin, R. L. and Chen, Ruyi and Lu, Shanghao and Zhou, Shangyan and Chen, Shanhuang and Ye, Shengfeng and Wang, Shiyu and Yu, Shuiping and Zhou, Shunfeng and Pan, Shuting and Li, S. S. and Zhou, Shuang and Wu, Shaoqing and Yun, Tao and Pei, Tian and Sun, Tianyu and Wang, T. and Zeng, Wangding and Liu, Wen and Liang, Wenfeng and Gao, Wenjun and Yu, Wenqin and Zhang, Wentao and Xiao, W. L. and An, Wei and Liu, Xiaodong and Wang, Xiaohan and Chen, Xiaokang and Nie, Xiaotao and Cheng, Xin and Liu, Xin and Xie, Xin and Liu, Xingchao and Yang, Xinyu and Li, Xinyuan and Su, Xuecheng and Lin, Xuheng and Li, X. Q. and Jin, Xiangyue and Shen, Xiaojin and Chen, Xiaosha and Sun, Xiaowen and Wang, Xiaoxiang and Song, Xinnan and Zhou, Xinyi and Wang, Xianzu and Shan, Xinxia and Li, Y. K. and Wang, Y. Q. and Wei, Y. X. and Zhang, Yang and Xu, Yanhong and Li, Yao and Zhao, Yao and Sun, Yaofeng and Wang, Yaohui and Yu, Yi and Zhang, Yichao and Shi, Yifan and Xiong, Yiliang and He, Ying and Piao, Yishi and Wang, Yisong and Tan, Yixuan and Ma, Yiyang and Liu, Yiyuan and Guo, Yongqiang and Ou, Yuan and Wang, Yuduan and Gong, Yue and Zou, Yuheng and He, Yujia and Xiong, Yunfan and Luo, Yuxiang and You, Yuxiang and Liu, Yuxuan and Zhou, Yuyang and Zhu, Y. X. and Huang, Yanping and Li, Yaohui and Zheng, Yi and Zhu, Yuchen and Ma, Yunxian and Tang, Ying and Zha, Yukun and Yan, Yuting and Ren, Z. Z. and Ren, Zehui and Sha, Zhangli and Fu, Zhe and Xu, Zhean and Xie, Zhenda and Zhang, Zhengyan and Hao, Zhewen and Ma, Zhicheng and Yan, Zhigang and Wu, Zhiyu and Gu, Zihui and Zhu, Zijia and Liu, Zijun and Li, Zilin and Xie, Ziwei and Song, Ziyang and Pan, Zizheng and Huang, Zhen and Xu, Zhipeng and Zhang, Zhongyu and Zhang, Zhen},
  author         = {Guo, Daya and Yang, Dejian and Zhang, Haowei and others},
  title          = {{DeepSeek-R1} incentivizes reasoning in LLMs through reinforcement learning},
  journal        = {Nature},
  year           = {2025},
  month          = {9},
  day            = {01},
  volume         = {645},
  number         = {8081},
  pages          = {633-638},
  issn           = {1476-4687},
  doi            = {10.1038/s41586-025-09422-z},
  comment-url    = {https://doi.org/10.1038/s41586-025-09422-z}
}

@article{designs4030031,
  author         = {Vivekananda, Ashish Alape and Enoiu, Eduard},
  title          = {Automated Test Case Generation for Digital System Designs: A Mapping Study on VHDL, Verilog, and SystemVerilog Description Languages},
  journal        = {Designs},
  volume         = {4},
  year           = {2020},
  number         = {3},
  article-number = {31},
  comment-url    = {https://www.mdpi.com/2411-9660/4/3/31},
  issn           = {2411-9660},
  doi            = {10.3390/designs4030031}
}

@article{electronics12183821,
  author         = {Liu, Cong and Xu, Xinyu and Chen, Zhenjiao and Wang, Binghao},
  title          = {A Universal-Verification-Methodology-Based Testbench for the Coverage-Driven Functional Verification of an Instruction Cache Controller},
  journal        = {Electronics},
  comment-volume = {12},
  year           = {2023},
  number         = {18},
  article-number = {3821},
  comment-url    = {https://www.mdpi.com/2079-9292/12/18/3821},
  issn           = {2079-9292},
  doi            = {10.3390/electronics12183821}
}

@misc{VeriPrefer,
  title         = {Insights from Verification: Training a Verilog Generation LLM with Reinforcement Learning with Testbench Feedback},
  author        = {Ning Wang and Bingkun Yao and Jie Zhou and Yuchen Hu and Xi Wang and Nan Guan and Zhe Jiang},
  year          = {2025},
  eprint        = {2504.15804},
  archiveprefix = {arXiv},
  primaryclass  = {cs.AR},
  comment-url   = {https://arxiv.org/abs/2504.15804}
}

@misc{yu2025dapoopensourcellmreinforcement,
  title          = {{DAPO}: An Open-Source LLM Reinforcement Learning System at Scale},
  comment-author = {Qiying Yu and Zheng Zhang and Ruofei Zhu and Yufeng Yuan and Xiaochen Zuo and Yu Yue and Weinan Dai and Tiantian Fan and Gaohong Liu and Lingjun Liu and Xin Liu and Haibin Lin and Zhiqi Lin and Bole Ma and Guangming Sheng and Yuxuan Tong and Chi Zhang and Mofan Zhang and Wang Zhang and Hang Zhu and Jinhua Zhu and Jiaze Chen and Jiangjie Chen and Chengyi Wang and Hongli Yu and Yuxuan Song and Xiangpeng Wei and Hao Zhou and Jingjing Liu and Wei-Ying Ma and Ya-Qin Zhang and Lin Yan and Mu Qiao and Yonghui Wu and Mingxuan Wang},
  author         = {Qiying Yu and Zheng Zhang and Ruofei Zhu and others},
  year           = {2025},
  eprint         = {2503.14476},
  archiveprefix  = {arXiv},
  primaryclass   = {cs.LG},
  comment-url    = {https://arxiv.org/abs/2503.14476}
}

@misc{pinckney2025revisitingverilogevalyearimprovements,
  title         = {Revisiting {VerilogEval}: A Year of Improvements in Large-Language Models for Hardware Code Generation},
  author        = {Nathaniel Pinckney and Christopher Batten and Mingjie Liu and Haoxing Ren and Brucek Khailany},
  year          = {2025},
  eprint        = {2408.11053},
  archiveprefix = {arXiv},
  primaryclass  = {cs.AR},
  comment-url   = {https://arxiv.org/abs/2408.11053}
}

@misc{Qwen,
  title          = {{Qwen2.5-Coder} Technical Report},
  comment-author = {Binyuan Hui and Jian Yang and Zeyu Cui and Jiaxi Yang and Dayiheng Liu and Lei Zhang and Tianyu Liu and Jiajun Zhang and Bowen Yu and Keming Lu and Kai Dang and Yang Fan and Yichang Zhang and An Yang and Rui Men and Fei Huang and Bo Zheng and Yibo Miao and Shanghaoran Quan and Yunlong Feng and Xingzhang Ren and Xuancheng Ren and Jingren Zhou and Junyang Lin},
  author         = {Binyuan Hui and Jian Yang and Zeyu Cui and others},
  year           = {2024},
  eprint         = {2409.12186},
  archiveprefix  = {arXiv},
  primaryclass   = {cs.CL},
  comment-url    = {https://arxiv.org/abs/2409.12186}
}

@misc{GPT4,
  title          = {{GPT-4} Technical Report},
  comment-author = {OpenAI and Josh Achiam and Steven Adler and Sandhini Agarwal and Lama Ahmad and Ilge Akkaya and Florencia Leoni Aleman and Diogo Almeida and Janko Altenschmidt and Sam Altman and Shyamal Anadkat and Red Avila and Igor Babuschkin and Suchir Balaji and Valerie Balcom and Paul Baltescu and Haiming Bao and Mohammad Bavarian and Jeff Belgum and Irwan Bello and Jake Berdine and Gabriel Bernadett-Shapiro and Christopher Berner and Lenny Bogdonoff and Oleg Boiko and Madelaine Boyd and Anna-Luisa Brakman and Greg Brockman and Tim Brooks and Miles Brundage and Kevin Button and Trevor Cai and Rosie Campbell and Andrew Cann and Brittany Carey and Chelsea Carlson and Rory Carmichael and Brooke Chan and Che Chang and Fotis Chantzis and Derek Chen and Sully Chen and Ruby Chen and Jason Chen and Mark Chen and Ben Chess and Chester Cho and Casey Chu and Hyung Won Chung and Dave Cummings and Jeremiah Currier and Yunxing Dai and Cory Decareaux and Thomas Degry and Noah Deutsch and Damien Deville and Arka Dhar and David Dohan and Steve Dowling and Sheila Dunning and Adrien Ecoffet and Atty Eleti and Tyna Eloundou and David Farhi and Liam Fedus and Niko Felix and Simón Posada Fishman and Juston Forte and Isabella Fulford and Leo Gao and Elie Georges and Christian Gibson and Vik Goel and Tarun Gogineni and Gabriel Goh and Rapha Gontijo-Lopes and Jonathan Gordon and Morgan Grafstein and Scott Gray and Ryan Greene and Joshua Gross and Shixiang Shane Gu and Yufei Guo and Chris Hallacy and Jesse Han and Jeff Harris and Yuchen He and Mike Heaton and Johannes Heidecke and Chris Hesse and Alan Hickey and Wade Hickey and Peter Hoeschele and Brandon Houghton and Kenny Hsu and Shengli Hu and Xin Hu and Joost Huizinga and Shantanu Jain and Shawn Jain and Joanne Jang and Angela Jiang and Roger Jiang and Haozhun Jin and Denny Jin and Shino Jomoto and Billie Jonn and Heewoo Jun and Tomer Kaftan and Łukasz Kaiser and Ali Kamali and Ingmar Kanitscheider and Nitish Shirish Keskar and Tabarak Khan and Logan Kilpatrick and Jong Wook Kim and Christina Kim and Yongjik Kim and Jan Hendrik Kirchner and Jamie Kiros and Matt Knight and Daniel Kokotajlo and Łukasz Kondraciuk and Andrew Kondrich and Aris Konstantinidis and Kyle Kosic and Gretchen Krueger and Vishal Kuo and Michael Lampe and Ikai Lan and Teddy Lee and Jan Leike and Jade Leung and Daniel Levy and Chak Ming Li and Rachel Lim and Molly Lin and Stephanie Lin and Mateusz Litwin and Theresa Lopez and Ryan Lowe and Patricia Lue and Anna Makanju and Kim Malfacini and Sam Manning and Todor Markov and Yaniv Markovski and Bianca Martin and Katie Mayer and Andrew Mayne and Bob McGrew and Scott Mayer McKinney and Christine McLeavey and Paul McMillan and Jake McNeil and David Medina and Aalok Mehta and Jacob Menick and Luke Metz and Andrey Mishchenko and Pamela Mishkin and Vinnie Monaco and Evan Morikawa and Daniel Mossing and Tong Mu and Mira Murati and Oleg Murk and David Mély and Ashvin Nair and Reiichiro Nakano and Rajeev Nayak and Arvind Neelakantan and Richard Ngo and Hyeonwoo Noh and Long Ouyang and Cullen O'Keefe and Jakub Pachocki and Alex Paino and Joe Palermo and Ashley Pantuliano and Giambattista Parascandolo and Joel Parish and Emy Parparita and Alex Passos and Mikhail Pavlov and Andrew Peng and Adam Perelman and Filipe de Avila Belbute Peres and Michael Petrov and Henrique Ponde de Oliveira Pinto and Michael and Pokorny and Michelle Pokrass and Vitchyr H. Pong and Tolly Powell and Alethea Power and Boris Power and Elizabeth Proehl and Raul Puri and Alec Radford and Jack Rae and Aditya Ramesh and Cameron Raymond and Francis Real and Kendra Rimbach and Carl Ross and Bob Rotsted and Henri Roussez and Nick Ryder and Mario Saltarelli and Ted Sanders and Shibani Santurkar and Girish Sastry and Heather Schmidt and David Schnurr and John Schulman and Daniel Selsam and Kyla Sheppard and Toki Sherbakov and Jessica Shieh and Sarah Shoker and Pranav Shyam and Szymon Sidor and Eric Sigler and Maddie Simens and Jordan Sitkin and Katarina Slama and Ian Sohl and Benjamin Sokolowsky and Yang Song and Natalie Staudacher and Felipe Petroski Such and Natalie Summers and Ilya Sutskever and Jie Tang and Nikolas Tezak and Madeleine B. Thompson and Phil Tillet and Amin Tootoonchian and Elizabeth Tseng and Preston Tuggle and Nick Turley and Jerry Tworek and Juan Felipe Cerón Uribe and Andrea Vallone and Arun Vijayvergiya and Chelsea Voss and Carroll Wainwright and Justin Jay Wang and Alvin Wang and Ben Wang and Jonathan Ward and Jason Wei and CJ Weinmann and Akila Welihinda and Peter Welinder and Jiayi Weng and Lilian Weng and Matt Wiethoff and Dave Willner and Clemens Winter and Samuel Wolrich and Hannah Wong and Lauren Workman and Sherwin Wu and Jeff Wu and Michael Wu and Kai Xiao and Tao Xu and Sarah Yoo and Kevin Yu and Qiming Yuan and Wojciech Zaremba and Rowan Zellers and Chong Zhang and Marvin Zhang and Shengjia Zhao and Tianhao Zheng and Juntang Zhuang and William Zhuk and Barret Zoph},
  author         = {OpenAI and Josh Achiam and Steven Adler and Sandhini Agarwal and others},
  year           = {2024},
  eprint         = {2303.08774},
  archiveprefix  = {arXiv},
  primaryclass   = {cs.CL},
  comment-url    = {https://arxiv.org/abs/2303.08774}
}

@misc{VeriReason,
  title         = {{VeriReason}: Reinforcement Learning with Testbench Feedback for Reasoning-Enhanced Verilog Generation},
  author        = {Yiting Wang and Guoheng Sun and Wanghao Ye and Gang Qu and Ang Li},
  year          = {2025},
  eprint        = {2505.11849},
  archiveprefix = {arXiv},
  primaryclass  = {cs.AI},
  comment-url   = {https://arxiv.org/abs/2505.11849}
}

\end{document}